\documentclass[fleqn,10pt]{wlscirep}

\usepackage{dcolumn}
\usepackage[utf8]{inputenc}
\usepackage[T1]{fontenc}
\usepackage{mathptmx}
\usepackage{graphicx}
\usepackage{bm}
\usepackage{fancyhdr}
\usepackage{bbold}
\usepackage{comment}
\usepackage{enumitem}
\usepackage{setspace}
\usepackage{makecell}
\usepackage{multirow}
\usepackage{appendix}
\setcellgapes{4pt}
\makegapedcells
\title{Radical pairs may play a role in xenon-induced general anesthesia}

\author[1,*]{Jordan Smith}
\author[1]{Hadi Zadeh Haghighi}
\author[2]{Dennis Salahub}
\author[1,*]{Christoph Simon}
\affil[1]{Department of Physics and Astronomy, Institute for Quantum Science and Technology, Quantum Alberta, and Hotchkiss Brain Institute, University of Calgary, Calgary, AB, T2N 1N4, Canada}
\affil[2]{Department of Chemistry, Department of Physics and Astronomy, Institute for Quantum Science and Technology, Quantum Alberta, Centre for Molecular Simulation, University of Calgary, Calgary, AB, T2N 1N4, Canada}

\affil[*]{jordan.smith1@ucalgary.ca, csimo@ucalgary.ca}

\begin{abstract}

Understanding the mechanisms underlying general anesthesia would be a key step towards understanding consciousness. The process of xenon-induced general anesthesia has been shown to involve electron transfer, and the potency of xenon as a general anesthetic exhibits isotopic dependence. We propose that these observations can be explained by a mechanism in which the xenon nuclear spin influences the recombination dynamics of a naturally occurring radical pair of electrons. We develop a simple model inspired by the body of work on the radical-pair mechanism in cryptochrome in the context of avian magnetoreception, and we show that our model can reproduce the observed isotopic dependence of the general anesthetic potency of xenon in mice. Our results are consistent with the idea that radical pairs of electrons with entangled spins could be important for consciousness. 

\end{abstract}

\begin{document}

\flushbottom
\maketitle

\newcommand{\beginsupplement}{%
        \setcounter{table}{0}
        \renewcommand{\thetable}{S\arabic{table}}%
        \setcounter{figure}{0}
        \renewcommand{\thefigure}{S\arabic{figure}}%
     }

\section*{\label{sec:Intro}Introduction}

Understanding consciousness remains one of the big open questions in neuroscience \cite{koch2016neural}, and in science in general.  
The study of anesthesia is one of the key approaches to elucidating the processes underlying consciousness \cite{gen_an2,mashour2006integrating}, but there are still significant open questions regarding the physical mechanisms of anesthesia itself \cite{franks2008general}.

  
  

One anesthetic agent that has been studied extensively is xenon. Xenon has been shown experimentally to produce a state of general anesthesia in several species, including \textit{Drosophila} \cite{xenon_fly}, mice \cite{xenon_mice}, and humans \cite{xenon_human}. While the anesthetic properties of xenon were discovered in 1939 \cite{xenon_anes}, the exact underlying mechanism by which it produces anesthetic effects remains unclear even after decades of research \cite{xenon_mech}. Our focus here is on this underlying physical mechanism, and there are important hints provided by two recent publications. 


First, Turin \textit{et al.} \cite{xenon_fly} showed that when xenon acts anesthetically on \textit{Drosophila} specific electron spin resonance (ESR) signals, consistent with free electrons, can be observed. Based on this observation, Turin \textit{et al.} \cite{xenon_fly} proposed that the anesthetic action of xenon may involve some form of electron transfer. They supported their proposal by density-functional theory (DFT) calculations showing the effect of xenon on nearby molecular orbitals. Second, Li \textit{et al.} \cite{xenon_mice} showed experimentally that isotopes of xenon with non-zero nuclear spin had reduced anesthetic potency in mice compared with isotopes with no nuclear spin. 

If the process by which xenon produces anesthetic effects includes free-electron transfer as well as nuclear-spin dependence, a mechanistic framework proposed to explain xenon-induced general anesthesia should possess these characteristics. Here we show that a model involving a radical pair of electrons (RP) and the subsequent modulation of the RP spin dynamics by hyperfine interactions (HFIs) is consistent with these assumptions.


The radical pair mechanism (RPM) was first proposed more than 50 years ago \cite{rpm3}. The rupture of a chemical bond can create a pair of electrons, where the electrons are localized on two different molecular entities, whose spins are entangled in a singlet state \cite{singlet}. The magnetic dipole moment associated with electron spin can interact and couple with other magnetic dipoles and external magnetic fields, including HFIs and Zeeman interactions \cite{rpm4}. As a consequence of such interactions, the initial singlet state can evolve into a more complex state that has both singlet and triplet components.  Eventually, the coherent oscillation of the RP between singlet and triplet states ceases and the electrons may recombine (for the singlet component of the state) or diffuse apart to form various triplet products \cite{rpm2}. The coherent spin dynamics and spin-dependent reactivity of RPs allow magnetic interactions which are six orders of magnitude smaller than the thermal energy, $k_BT$, to have predictable and reproducible effects on chemical reaction yields \cite{rpm5,Crypt}.


The RPM has become a prominent concept in quantum biology \cite{lambert2013quantum}. In particular, it has been studied in detail for the cryptochrome protein as a potential explanation for avian magnetoreception \cite{hore2016radical,Aviation,crypt2,photo,rpm5,alt,Aviation2}. In the present work we apply the principles and methods used to investigate cryptochrome \cite{Crypt} to xenon-induced general anesthesia.

General anesthetics produce widespread neurodepression in the central nervous system by enhancing inhibitory neurotransmission and reducing excitatory neurotransmission across synapses \cite{ion_channel,nmda}. Three ligand-gated ion-channels in particular have emerged as likely molecular targets for a range of anesthetic agents \cite{Glycine2}: the inhibitory glycine receptor, the inhibitory $\gamma$-aminobutyric acid type-A (GABA$_A$) receptor, and the excitatory N-methyl-D-aspartate (NMDA) receptor \cite{xenon_how}. NMDA receptors require both glycine and glutamate for excitatory activation \cite{nmda2}, and it has been suggested that xenon's anesthetic action is related to xenon atoms participating in competitive inhibition of the NMDA receptor in central-nervous-system neurons by binding at the glycine binding site in the NMDA receptor \cite{Glycine2,xenon_how,Glycine,Xenon_med}. Armstrong \textit{et al.} \cite{Glycine2,Glycine} propose that the glycine binding site of the NMDA receptor contains aromatic residues such as tryptophan (Trp), similar to those found in cryptochrome, which is consistent with the possibility that the anesthetic action of xenon involves a spin-dependent process similar to the RPM thought to occur in cryptochrome. Let us note that other targets have also been proposed for many anesthetics, e.g. tubulin \cite{craddock2017anesthetic}. In the following we focus on the NMDA receptor to be specific, but our model is quite general and could well apply to other targets.


We suggest that the (possibly partial) electron transfer related to xenon's anesthetic action that is evidenced by Turin \textit{et al.} \cite{xenon_fly} plays a role in the recombination dynamics of a naturally occurring RP and that, for isotopes of xenon with a non-zero nuclear spin, this nuclear spin couples with (at least one of) the electron spins of such a RP, affecting the reaction yields of the RP and hence xenon's anesthetic action. Such a mechanism is consistent with the experimental results of Li \textit{et al.} \cite{xenon_mice} that xenon isotopes with non-zero nuclear spin have reduced anesthetic potency compared to isotopes with zero nuclear spin.

Aromatic residues are important for the binding of xenon and glycine at the glycine binding site of the NMDA receptor. While Armstrong \textit{et al.} \cite{Glycine} have shown that phenylalanine residues within the NMDA receptor are necessary for the binding of xenon in the active site, the redox inactivity of phenylalanine \cite{byrdin2007observation, li1991active} makes it unlikely to be involved in RP formation. Trp, however, is redox active, as evidenced by its participation in RP formation in the context of cryptochrome, and could feasibly be involved in the formation of a RP here. It has been suggested that water (a source of oxygen) may be present in the NMDA receptor \cite{Glycine2,nmda2}, and Aizenman \textit{et al.} \cite{oxygen2} as well as Girouard \textit{et al.} \cite{oxygen} suggest that reactive oxygen species (ROS) may be found within the NMDA receptor. Further, Turin and Skoulakis \cite{xenon_o2} found that when a sample of xenon gas was administered to \textit{Drosophila} without oxygen gas present in the sample, no spin changes were observed in the flies. Motivated by these observations, we propose that a Trp residue located in the glycine binding site of the NMDA receptor could be oxidized by a nearby ROS, forming a [O$_2^{\cdot -}$ TrpH$^{\cdot +}$] RP. We consider a scenario in which a xenon atom interacts with the O$_2^{\cdot -}$ radical electron, shifting some of the electron density from the oxygen onto the xenon atom (see also Methods, Fig. \ref{dft_homo}), which results in HFI between the xenon nuclear spin and the O$_2^{\cdot -}$ radical electron spin. The xenon nuclear spin is consequently able to influence the spin-dynamics and ultimate product yields of the [O$_2^{\cdot -}$ TrpH$^{\cdot +}$] RP, provided that the hyperfine interaction is sufficiently strong compared to the spin relaxation and reaction rates. Similarly to our above remark on the target receptor, here we focus on one specific potential radical pair for concreteness, but other implementations of the RPM may also be viable explanations for the mentioned experimental observations.

Given that cryptochrome is one case in which the RPM has been studied extensively, we propose that by recognizing commonalities in the biological and chemical environments in which magnetoreception and xenon-induced general anesthesia are thought to take place, analytical and numerical techniques that have been used to study cryptochrome may be adapted and applied to the case of xenon-induced anesthesia, potentially providing insight into general anesthetic mechanisms. We explore the feasibility of such a mechanism by determining and analyzing the necessary parameters and conditions under which the spin-dependent RP product yields can explain the experimental isotope-dependent anesthetic effects reported by Li \textit{et al} \cite{xenon_mice}.

\section*{\label{sec:Res}Results}

\subsection*{\label{subsec:xenon_spin_dep}Predicting Experimental Xenon Anesthesia Results using the RPM Model}

\subsubsection*{\label{subsec:xenon_pot}Quantifying experimental anesthetic potency}

In the work of Li \textit{et al.} \cite{xenon_mice}, a metric referred to as the ``loss of righting reflex ED50'' (LRR-ED50) was defined using the concentration of xenon administered to mice, in which the mice were no longer able to right themselves within 10 s of being flipped onto their backs. The LRR-ED50 metric was reported to be correlated with consciousness in mice, and was measured experimentally for $^{132}$Xe, $^{134}$Xe, $^{131}$Xe, and $^{129}$Xe to be 70(4)\%, 72(5)\%, 99(5)\%, and 105(7)\%, respectively \cite{xenon_mice}.

Here we define the anesthetic potency as the inverse of the LRR-ED50 metric. In order to quantify the relative anesthetic potency of the various xenon isotopes, the potency of $^{132}$Xe (with $I=0$) is normalized to 1. The inverse of the LRR-ED50 value of isotopes $^{131}$Xe and $^{129}$Xe is then divided by that of $^{132}$Xe. The relative isotopic anesthetic potencies are quantified as $Pot_{0}=1$, $Pot_{3/2}=0.71(8)$, and $Pot_{1/2}=0.67(8)$ for $^{132}$Xe, $^{131}$Xe, and $^{129}$Xe, respectively, as shown in Table \ref{isotopes}, where $Pot_{0}$ is the relative potency of xenon with nuclear spin $I=0$, and likewise for $Pot_{3/2}$ and $Pot_{1/2}$.

\begin{table}[ht!]
\caption{Xenon isotopic nuclear spin, LRR-ED50, and $Pot$ values for xenon isotopes $^{132}$Xe, $^{134}$Xe, $^{131}$Xe, and $^{129}$Xe. LRR-ED50 values as reported in the work of Li \textit{et al.} \cite{xenon_mice}}
\label{isotopes}
\centering
\begin{tabular}{|c|c|c|c|}
\hline
Isotope    & Nuclear Spin, $I$ & LRR-ED50 (\%) & $Pot$   \\ \hline
$^{132}$Xe & 0                 & 70(4)         & 1       \\ \hline
$^{134}$Xe & 0                 & 72(5)         & -       \\ \hline
$^{131}$Xe & 3/2               & 99(5)         & 0.71(8) \\ \hline
$^{129}$Xe & 1/2               & 105(7)        & 0.67(8) \\ \hline
\end{tabular}
\end{table}

\subsubsection*{\label{subsubsec:RPM}RPM model}

\begin{figure}[ht!]
\centerline{
\includegraphics[width=0.30\textwidth,height=1.0\textheight,keepaspectratio]
{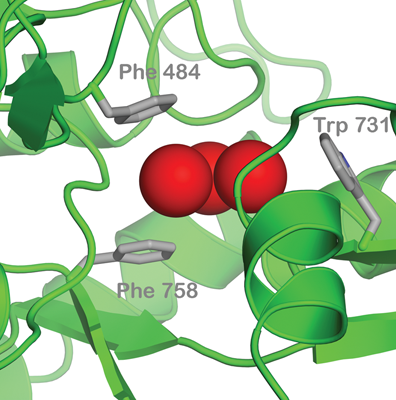}}
\caption{The image shows the predicted position of xenon atoms (red spheres) in the glycine site of the NMDA receptor, together with the aromatic residues phenylalanine 758, phenylalanine 484, and tryptophan 731. \cite{Glycine}}
\label{xenon_binding_site}
\end{figure}

We have developed a RPM model to predict anesthetic potency by making a connection with the relative singlet yield for different isotopes, as described in more detail below. The model that we have used here was developed using a hypothetical xenon-NMDA receptor RP system, based on the information about xenon action sites mentioned previously and shown in Fig. \ref{xenon_binding_site}. Our model was also developed by considering the cryptochrome case as related to magnetoreception. Here we have proposed that the xenon-NMDA receptor RP system may involve xenon, ROS, and Trp, located in the glycine-binding site of the NMDA receptor. A spin-correlated RP of electrons (termed electrons A and B), possibly in the form of [O$_2^{\cdot -}$ TrpH$^{\cdot +}$], interact with xenon, where the O$_2^{\cdot -}$ radical electron (electron A) spin couples with non-zero xenon nuclear spin. 


The number of xenon atoms located in the active site when anesthetic action takes place is not yet completely clear, and the work of Dickinson \textit{et al.} \cite{Glycine2} suggests that the number of xenon atoms simultaneously present in the active site ranges probabilistically between zero and three, with single-xenon occupation proposed to be most probable. 
Here we show that the simplest and most probable case of a single xenon atom occupying the active site allows us to explain the anesthetic potency ratios derived from the experimental results of Li \textit{et al.} \cite{xenon_mice} The additional degrees of freedom implicit in more complex Hilbert spaces, such as the cases of two and three-xenon occupation states, only aid the model in explaining the experimental results of Li \textit{et al} \cite{xenon_mice}. We therefore focus our analysis on the single-xenon case. 

The Hamiltonian of the RPM in the case of xenon-induced anesthesia depends not only upon Zeeman interactions and the number of xenon atoms present in the glycine binding site of the NMDA receptor, but also upon the assumed HFIs. Due to the likely randomized orientation of the relevant receptors in the brain, we consider only the isotropic Fermi contact contribution to the HFIs. In the simulation involving a single xenon atom occupying the active site, it is assumed that the O$_2^{\cdot -}$ radical electron couples only to the xenon nucleus. Most of our calculations below are based on the simplifying assumption that the TrpH$^{\cdot +}$ electron (electron B) couples only to the nuclear spin of the indole nitrogen in TrpH$^{\cdot +}$, following the treatment of Hore \cite{Crypt}, but we also briefly consider a separate scenario in which electron B couples only to the nuclear spin of the TrpH$^{\cdot +}$ $\beta$-hydrogen, which we select because it has the largest isotropic hyperfine coupling constant of all 14 nuclear spins in tryptophan according to Maeda \textit{et al.} \cite{maeda2012}. The Hamiltonian describing these interactions is given as
\begin{equation}
\hat{H} = \omega \left( \hat{S}_{Az} + \hat{S}_{Bz} \right) + a_1 \hat{\textbf{S}}_{A} \cdot \hat{\textbf{I}}_{1} + a_2 \hat{\textbf{S}}_{B} \cdot \hat{\textbf{I}}_{2},
\label{hamiltonian1}
\end{equation}
\noindent where $\hat{\textbf{S}}_{A}$ and $\hat{\textbf{S}}_{B}$ are the spin operators of radical electrons A and B, respectively, $\hat{\textbf{I}}_{1}$ is the nuclear spin operator of the xenon nucleus, $\hat{\textbf{I}}_{2}$ is the nuclear spin operator of the TrpH$^{\cdot +}$ residue (where $I = 1$ for the indole nitrogen, and $I = 1/2$ for the $\beta$-hydrogen), $a_2 = \gamma_e a_2'$ is the hyperfine coupling constant describing the strength of the HFI between radical electron B and the nuclear spin in TrpH$^{\cdot +}$ ($a_2' = 347\ \mu$T for the indole N, and $a_2' = 1600\ \mu$T for the $\beta$-hydrogen \cite{maeda2012}), $a_1 = (\gamma_n / |\gamma_{^{129}\mbox{Xe}}|) \gamma_e a_1'$ is the hyperfine coupling constant describing the HFI strength between radical electron A and the xenon nucleus, where $\gamma_n$ is the nuclear gyromagnetic ratio of a given xenon isotope, $\gamma_{^{129}\mbox{Xe}}$ is the nuclear gyromagnetic ratio of $^{129}$Xe, $\gamma_e$ is the gyromagnetic ratio of an electron, and $\omega$ is the Larmor precession frequency of the electrons about an external magnetic field \cite{Crypt}. The Larmor precession frequency is defined as  $\omega = \gamma_e B$, where $B$ is the external magnetic field strength (assumed to be along the z-axis). In the calculation of $a_1$, the correction factor of $(\gamma_n / |\gamma_{^{129}\mbox{Xe}}|)$ is included to account for the unique nuclear gyromagnetic ratios of the various xenon isotopes, where the magnitude of this correction factor is normalized to 1 for $^{129}$Xe. The isotopic xenon gyromagnetic ratios are taken to be \cite{joki2017} $\gamma_{^{129}\mbox{Xe}} = -7.441\times10^7$ rad(sT)$^{-1}$, $\gamma_{^{131}\mbox{Xe}} = 2.206\times10^7$ rad(sT)$^{-1}$, and $\gamma_{^{132}\mbox{Xe}} = 0$. Consequently, the hyperfine coupling constant $a_1'$ is multiplied by factors $[-1,\ (2.206/7.441),\ 0]$ when calculated for each respective isotope. It should be pointed out that we focus here on the HFIs previously described, and that interactions between the two electron spins, as well as potential interactions between the electron spins and other nuclei are neglected. 

\subsubsection*{\label{subsubsec:DFT}Using DFT to calculate Hyperfine Coupling Constants}

As a means of checking if the prediction results of our xenon-induced anesthesia model are reasonable, we use DFT to study the interaction between O$_2^{\cdot -}$ and xenon, and to calculate the value of the expected hyperfine coupling constant $a_{1, exp}'$. Depending on the functional and basis-set used, we find a range of values $a_{1, exp}' \in [955,\ 2644]\ \mu$T.  We also use DFT analysis to calculate a value of $a_{2, exp}' = 347\ \mu$T for the hyperfine coupling between the radical electron on TrpH$^{\cdot +}$ and the nitrogen nuclear spin. Our value for the hyperfine coupling between the radical electron on TrpH$^{\cdot +}$ and the $\beta$-hydgrogen nuclear spin is taken from Maeda {\it et al.} \cite{maeda2012}. More details about our DFT analysis can be found in Methods.

\subsubsection*{\label{subsubsec:singlet_yield}Determination of Singlet Yield Ratios}

The eigenvalues and eigenvectors of the Hamiltonian can be used to determine the ultimate singlet yield ($\Phi_S$) for all times much greater than the radical-pair lifetime ($t \gg \tau$): \cite{Crypt}
\begin{equation}
\Phi_S = \frac{1}{4} - \frac{k}{4(k + r)} + \frac{1}{M} \sum^{4M}_{m=1} \sum^{4M}_{n=1} \left\lvert \left\langle m \right\rvert \hat{P}^S \left\lvert n \right\rangle \right\rvert ^2 \frac{k(k + r)}{(k + r)^2 + (\omega_m - \omega_n)^2},
\label{yield}
\end{equation}

\noindent where, following the methodology used by Hore \cite{Crypt} in the context of cryptochrome, $M$ is the total number of nuclear spin configurations, $\hat{P}^S$ is the singlet projection operator, $\left\lvert m \right\rangle$ and $\left\lvert n \right\rangle$ are eigenstates of $\hat{H}$ with corresponding energies of $\omega_m = \left\langle m \right\rvert \hat{H} \left\lvert m \right\rangle$ and $\omega_n = \left\langle n \right\rvert \hat{H} \left\lvert n \right\rangle$, respectively, $k=\tau^{-1}$ is  inverse of the RP lifetime, and $r=\tau_c^{-1}$ is the inverse of the RP spin-coherence lifetime.

The spin-dependent RP singlet yield is calculated for each xenon isotope under consideration. The singlet yield ratio ($SR$) for each xenon nuclear spin value is then calculated by dividing the singlet yield obtained using the given spin value by the singlet yield obtained using spin $I=0$, resulting in the ratio of spin-1/2 singlet yield to spin-0 singlet yield being expressed as $SR_{1/2}$, and the ratio of spin-3/2 singlet yield to spin-0 singlet yield expressed as $SR_{3/2}$, with the singlet yield ratio of spin-0 being normalized to $SR_{0} = 1$. The calculated singlet yield ratios are compared with the xenon potency ratios ($Pot$) derived using the data reported by Li \textit{et al.} \cite{xenon_mice}, as described above.



\begin{figure}[ht!]
\centerline{
\includegraphics[width=1.05\textwidth,height=1.0\textheight,keepaspectratio]
{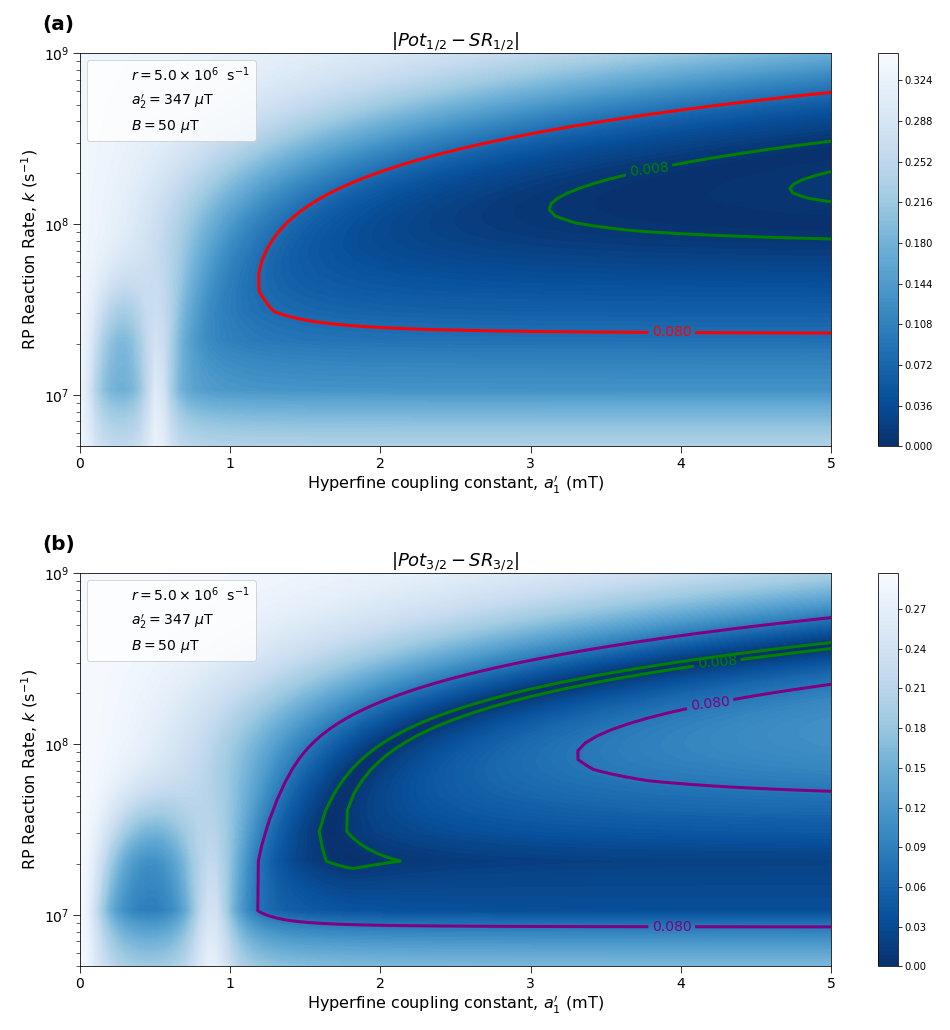}}
\caption{The dependence of the agreement between relative anesthetic potency and singlet yield ratio on changes in the hyperfine coupling constant $a_1'$ and the RP reaction rate $k$ for $a_1' \in [0,\ 5000]\ \mu$T and $k \in [5.0 \times 10^5,\ 1.0 \times 10^9]$ s$^{-1}$, using $a_2' = 347\ \mu$T, $B = 50\ \mu$T, and $r = 5.0 \times 10^{6}$ s$^{-1}$. The singlet yield ratio $SR$ is calculated using $a_1 = (\gamma_n / |\gamma_{^{129}\mbox{Xe}}|) \gamma_e a_1'$, where the magnitude of the isotopic nuclear gyromagnetic correction factor, $(\gamma_n / |\gamma_{^{129}\mbox{Xe}}|)$, is normalized for $^{129}$Xe. The model can explain the experimentally derived \cite{xenon_mice} relative anesthetic potency of xenon for values of $a_1'$ and $k$ where $\lvert Pot_{1/2} - SR_{1/2}\rvert,\ \lvert Pot_{3/2} - SR_{3/2}\rvert \leq 0.08$. \textbf{(a)} The absolute difference between $Pot_{1/2}$ and $SR_{1/2}$. \textbf{(b)} The absolute difference between $Pot_{3/2}$ and $SR_{3/2}$.}
\label{hyper_k_dep1}
\end{figure}
\begin{figure}[ht!]
\centerline{
\includegraphics[width=1.08\textwidth,height=1.0\textheight,keepaspectratio]
{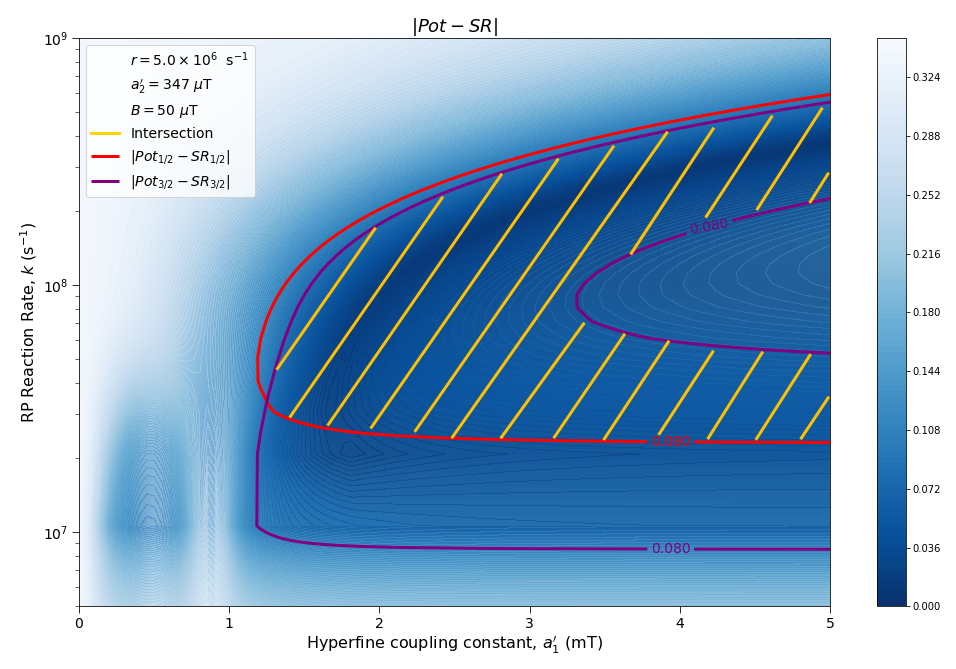}}
\caption{The dependence of the agreement between relative anesthetic potency and singlet yield ratio on changes in the hyperfine coupling constant $a_1'$ and the RP reaction rate $k$ for $a_1' \in [0,\ 5000]\ \mu$T and $k \in [5.0 \times 10^5,\ 1.0 \times 10^9]$ s$^{-1}$, using $a_2' = 347\ \mu$T, $B = 50\ \mu$T, and $r = 5.0 \times 10^{6}$ s$^{-1}$. The singlet yield ratio $SR$ is calculated using $a_1 = (\gamma_n / |\gamma_{^{129}\mbox{Xe}}|) \gamma_e a_1'$, where the magnitude of the isotopic nuclear gyromagnetic correction factor, $(\gamma_n / |\gamma_{^{129}\mbox{Xe}}|)$, is normalized for $^{129}$Xe. The model can explain the experimentally derived \cite{xenon_mice} relative anesthetic potency of xenon where $( \lvert Pot_{1/2} - SR_{1/2}\rvert \leq 0.08)$ and $(\lvert Pot_{3/2} - SR_{3/2}\rvert \leq 0.08)$ intersect, shaded in yellow.}
\label{hyper_k_dep_comb1}
\end{figure}

Here we analyze the sensitivity of the singlet yield ratios to changes in the HFI between radical electron A and the xenon nucleus ($a_1'$), RP reaction rate ($k$), RP spin-coherence relaxation rate ($r$), and external magnetic field strength ($B$). The dependence of the quantity $\lvert Pot - SR \rvert$ on $a_1'$ and $r$ is shown in Figs. \ref{hyper_r_dep1}/\ref{hyper_r_dep_comb1}, the dependence of $\lvert Pot - SR \rvert$ on $a_1'$ and $k$ is shown in Figs. \ref{hyper_k_dep1}/\ref{hyper_k_dep_comb1}, and the dependence of $\lvert Pot - SR \rvert$ on $r$ and $k$ is shown in Figs. \ref{rk_dep1}/\ref{rk_dep1_comb}. The relationship between $SR$ and $B$ can be seen in Fig. \ref{xenon_plot_extfield}. We also look at the dependence of the singlet yield ratios on the RP spin-coherence relaxation rate and the RP reaction rate in the case where electron B couples with the TrpH$^{\cdot +}$ $\beta$-hydrogen. The dependence of $\lvert Pot - SR \rvert$ on $r$ and $k$ in the electron B-$\beta$-hydrogen case can be seen in Fig. \ref{xenon_betahydrogen_rk}. 
Our goal is to find regions in parameter space such that the spin-dependent singlet yield ratios match the anesthetic isotopic potency ratios, i.e., the quantities $\lvert Pot_{1/2} - SR_{1/2}\rvert$ and $\lvert Pot_{3/2} - SR_{3/2}\rvert$ should be smaller than the experimental uncertainties in the anesthetic potency. 


The parameter space for which our model correctly predicts the experimentally derived  anesthetic potency results within uncertainty is quite broad for any given parameter, depending on the fixed values of the other parameters, and is found to be $a_1' \in [1207,\ 5000]\ \mu$T (normalized for $^{129}$Xe), $r \in [5.0 \times 10^{5},\ 1.7 \times 10^{8}]$ s$^{-1}$, $k \in [1.8 \times 10^{7},\ 5.5 \times 10^{8}]$ s$^{-1}$ for $a_2' = a_{2, exp}' = 347\ \mu$T and $B = 50\ \mu$T, where the searched parameter space includes $a_1' \in [0,\ 5000]\ \mu$T, and $r, k \in [5.0 \times 10^{5},\ 1.0 \times 10^{9}]$ s$^{-1}$. The parameter values that minimize the ``combined error'' $CE = |Pot_{1/2} - SR_{1/2}| + |Pot_{3/2} - SR_{3/2}|$ are found to be $a_1' = 1896\ \mu$T, $k = 8.6 \times 10^7$ s$^{-1}$, and $r = 1.0 \times 10^{6}$ s$^{-1}$, where $a_2' = a_{2, exp}'= 347\ \mu$T and $B = 50\ \mu$T are fixed, resulting in $SR_{1/2} = 0.70$ and $SR_{3/2} = 0.71$. It should be noted here that while $\lvert Pot_{3/2} - SR_{3/2} \rvert$ is less than $\lvert Pot_{1/2} - SR_{1/2} \rvert$ for these parameter values, both $SR_{3/2}$ and $SR_{1/2}$ are consistent with the experimental results of Li \textit{et al.} \cite{xenon_mice}, within their uncertainties. Further experiments with less uncertainty about the relative anesthetic potencies of spin-3/2 and spin-1/2 xenon isotopes would be of interest.

\section*{\label{sec:Dis}Discussion}

 
The main question that we set out to answer in this study was whether the RPM can provide a viable explanation for the isotope effects observed by Li {\it et al.} \cite{xenon_mice}. Our above results answer this question in the affirmative. A simple radical pair model with reasonable parameter values can reproduce the observed anesthetic potencies. 

To be concrete, we proposed a specific scenario involving a [O$_2^{\cdot -}$ TrpH$^{\cdot +}$] RP, where the oxygen molecule interacts with a single xenon atom. This proposal is motivated by the suggestions that Trp is present in the relevant binding site of the NMDA receptor \cite{Glycine} and that ROS may be found within the NMDA receptor \cite{oxygen2, oxygen}, as well as the observation that oxygen should be present when xenon is administered to \textit{Drosophila} in order to observe the ESR signals that correlate with xenon-induced general anesthesia \cite{xenon_o2}. The RP spin-coherence relaxation rate, RP reaction rate, and hyperfine coupling parameters were varied to reproduce the experimental results of Li \textit{et al.} \cite{xenon_mice}, and the resulting range of values of the hyperfine coupling constant $a_1'$ was found to agree with the range of values calculated for xenon in the proximity of O$_2^{\cdot -}$ using DFT analysis. 
 
\begin{figure}[ht!]
\centerline{
\includegraphics[width=1.05\textwidth,height=1.0\textheight,keepaspectratio]
{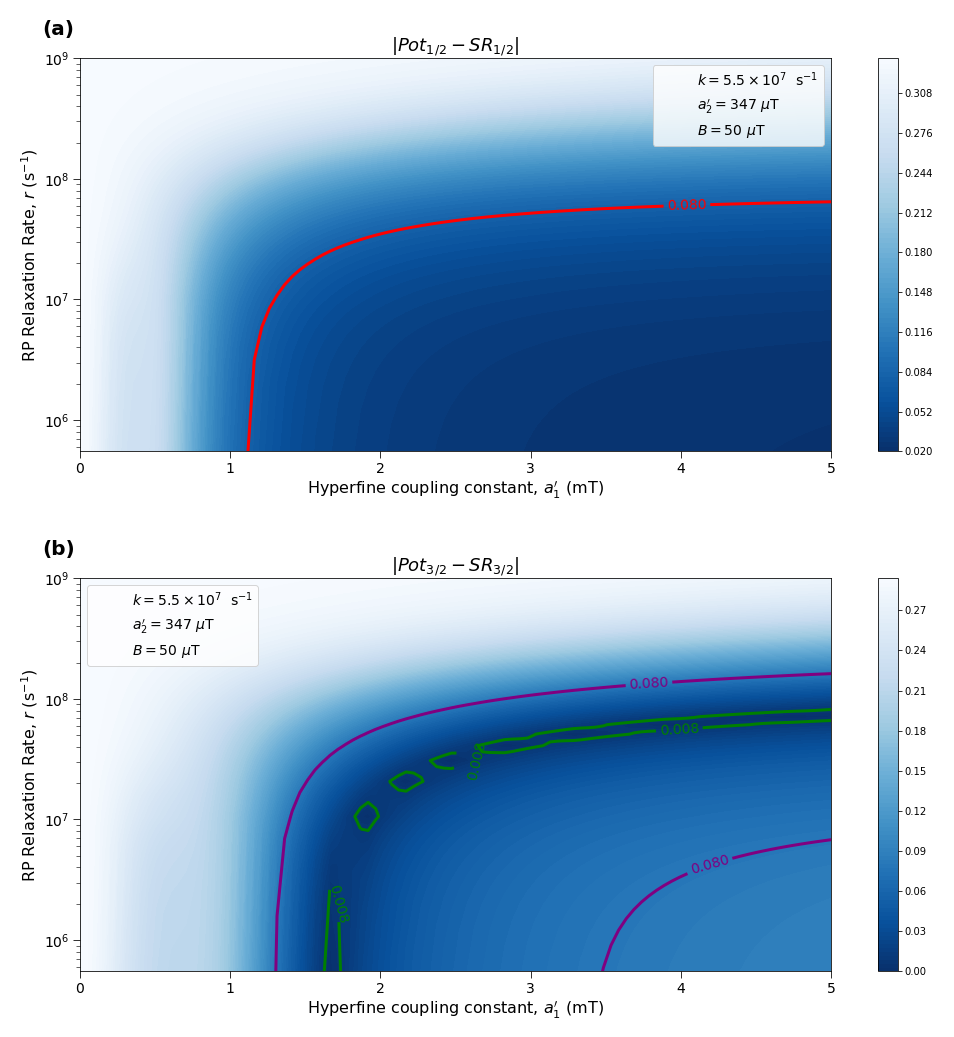}}
\caption{The dependence of the agreement between relative anesthetic potency and singlet yield ratio on changes in the hyperfine coupling constant $a_1'$ and the RP spin-coherence relaxation rate $r$ for $a_1' \in [0,\ 5000]\ \mu$T and $r \in [5.0 \times 10^5,\ 1.0 \times 10^9]$ s$^{-1}$, using $a_2' = 347\ \mu$T, $B = 50\ \mu$T, and $k = 5.5 \times 10^{7}$ s$^{-1}$. The singlet yield ratio $SR$ is calculated using $a_1 = (\gamma_n / |\gamma_{^{129}\mbox{Xe}}|) \gamma_e a_1'$, where the magnitude of the isotopic nuclear gyromagnetic correction factor, $(\gamma_n / |\gamma_{^{129}\mbox{Xe}}|)$, is normalized for $^{129}$Xe. The model can explain the experimentally derived \cite{xenon_mice} relative anesthetic potency of xenon for values of $a_1'$ and $r$ where $\lvert Pot_{1/2} - SR_{1/2}\rvert,\ \lvert Pot_{3/2} - SR_{3/2}\rvert \leq 0.08$. \textbf{(a)} The absolute difference between $Pot_{1/2}$ and $SR_{1/2}$. \textbf{(b)} The absolute difference between $Pot_{3/2}$ and $SR_{3/2}$.}
\label{hyper_r_dep1}
\end{figure}

\begin{figure}[ht!]
\centerline{
\includegraphics[width=1.08\textwidth,height=1.0\textheight,keepaspectratio]
{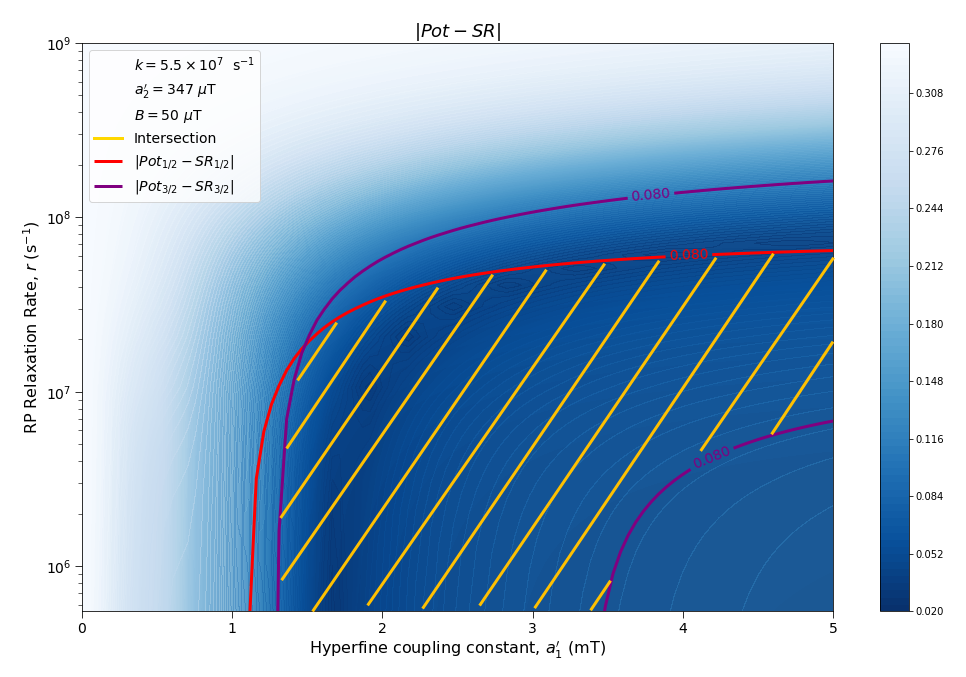}}
\caption{The dependence of the agreement between relative anesthetic potency and singlet yield ratio on changes in the hyperfine coupling constant $a_1'$ and the RP spin-coherence relaxation rate $r$ for $a_1' \in [0,\ 5000]\ \mu$T and $r \in [5.0 \times 10^5,\ 1.0 \times 10^9]$ s$^{-1}$, using $a_2' = 347\ \mu$T, $B = 50\ \mu$T, and $k = 5.5 \times 10^{7}$ s$^{-1}$. The singlet yield ratio $SR$ is calculated using $a_1 = (\gamma_n / |\gamma_{^{129}\mbox{Xe}}|) \gamma_e a_1'$, where the magnitude of the isotopic nuclear gyromagnetic correction factor, $(\gamma_n / |\gamma_{^{129}\mbox{Xe}}|)$, is normalized for $^{129}$Xe. The model can explain the experimentally derived \cite{xenon_mice} relative anesthetic potency of xenon where $( \lvert Pot_{1/2} - SR_{1/2}\rvert \leq 0.08)$ and $(\lvert Pot_{3/2} - SR_{3/2}\rvert \leq 0.08)$ intersect, shaded in yellow.}
\label{hyper_r_dep_comb1}
\end{figure}

\begin{figure}[ht!]
\centerline{
\includegraphics[width=1.05\textwidth,height=1.0\textheight,keepaspectratio]
{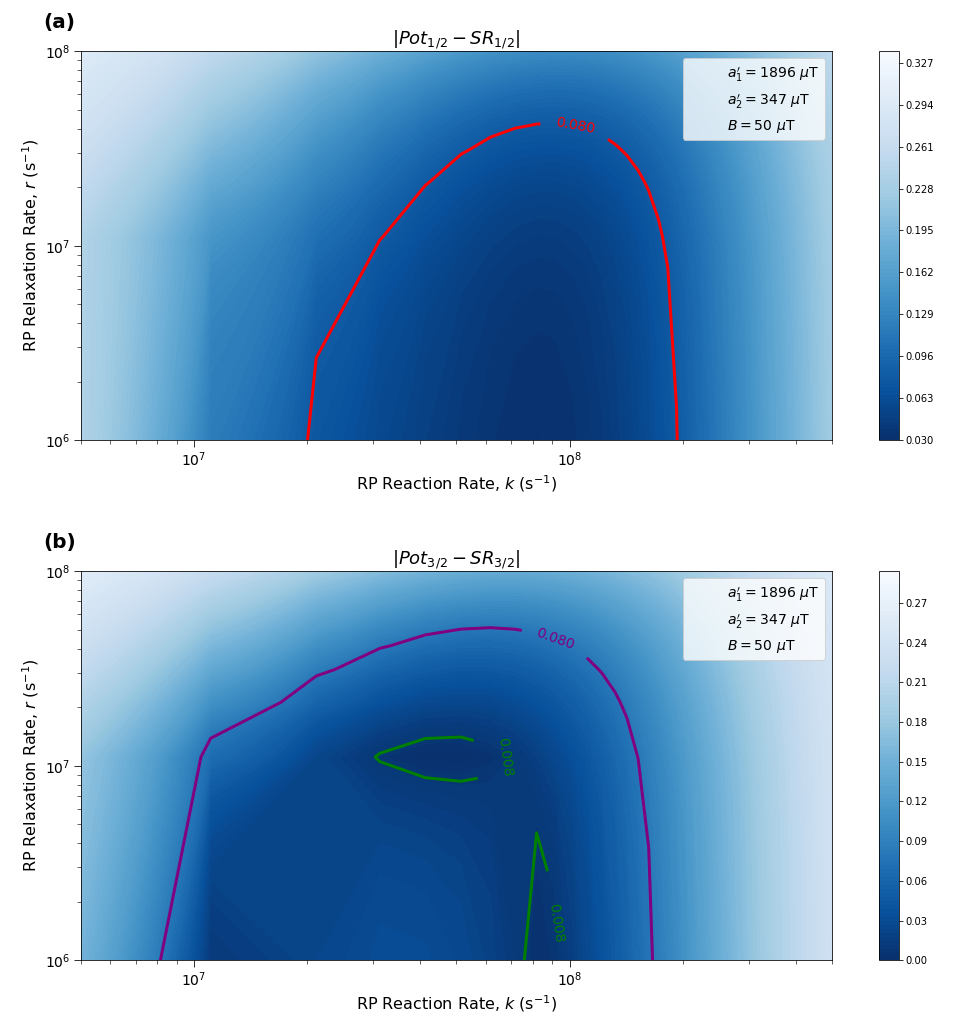}}
\caption{The dependence of the agreement between relative anesthetic potency and singlet yield ratio on the relationship between $r$ and $k$ for $r \in [1.0 \times 10^{6},\ 1.0 \times 10^{8}]$ s$^{-1}$ and $k \in [5.0 \times 10^6,\ 5.0 \times 10^8]$ s$^{-1}$, using $a_1' = 1896\ \mu$T, $a_2' = 347\ \mu$T, and $B = 50\ \mu$T. The singlet yield ratio $SR$ is calculated using $a_1 = (\gamma_n / |\gamma_{^{129}\mbox{Xe}}|) \gamma_e a_1'$, where the magnitude of the isotopic nuclear gyromagnetic correction factor, $(\gamma_n / |\gamma_{^{129}\mbox{Xe}}|)$, is normalized for $^{129}$Xe. The model can explain the experimentally derived \cite{xenon_mice} relative anesthetic potency of xenon for values of $r$ and $k$ where $\lvert Pot_{1/2} - SR_{1/2}\rvert,\ \lvert Pot_{3/2} - SR_{3/2}\rvert \leq 0.08$. \textbf{(a)} The absolute difference between $Pot_{1/2}$ and $SR_{1/2}$. \textbf{(b)} The absolute difference between $Pot_{3/2}$ and $SR_{3/2}$.}
\label{rk_dep1}
\end{figure}

\begin{figure}[ht!]
\centerline{
\includegraphics[width=1.08\textwidth,height=1.0\textheight,keepaspectratio]
{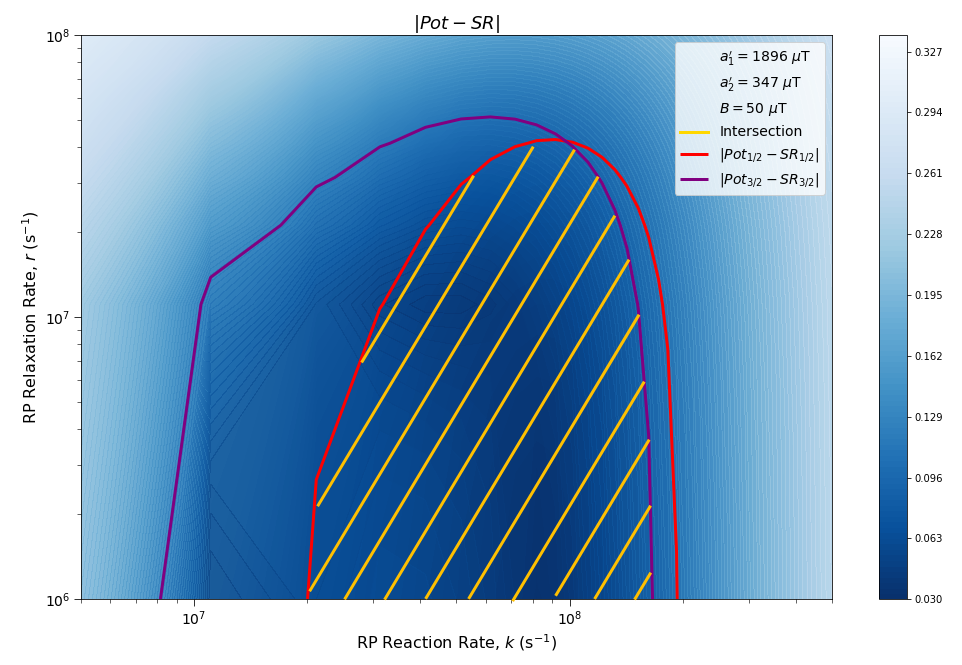}}
\caption{The dependence of the agreement between relative anesthetic potency and singlet yield ratio on the relationship between $r$ and $k$ for $r \in [1.0 \times 10^{6},\ 1.0 \times 10^{8}]$ s$^{-1}$ and $k \in [5.0 \times 10^6,\ 5.0 \times 10^8]$ s$^{-1}$, using $a_1' = 1896\ \mu$T, $a_2' = 347\ \mu$T, and $B = 50\ \mu$T. The singlet yield ratio $SR$ is calculated using $a_1 = (\gamma_n / |\gamma_{^{129}\mbox{Xe}}|) \gamma_e a_1'$, where the magnitude of the isotopic nuclear gyromagnetic correction factor, $(\gamma_n / |\gamma_{^{129}\mbox{Xe}}|)$, is normalized for $^{129}$Xe. The model can explain the experimentally derived \cite{xenon_mice} relative anesthetic potency of xenon where $\lvert Pot_{1/2} - SR_{1/2}\rvert \leq 0.08$ and $\lvert Pot_{3/2} - SR_{3/2}\rvert \leq 0.08$ intersect, shaded in yellow.}
\label{rk_dep1_comb}
\end{figure}

\begin{figure}[ht!]
\centerline{
\includegraphics[width=1.08\textwidth,height=1.0\textheight,keepaspectratio]
{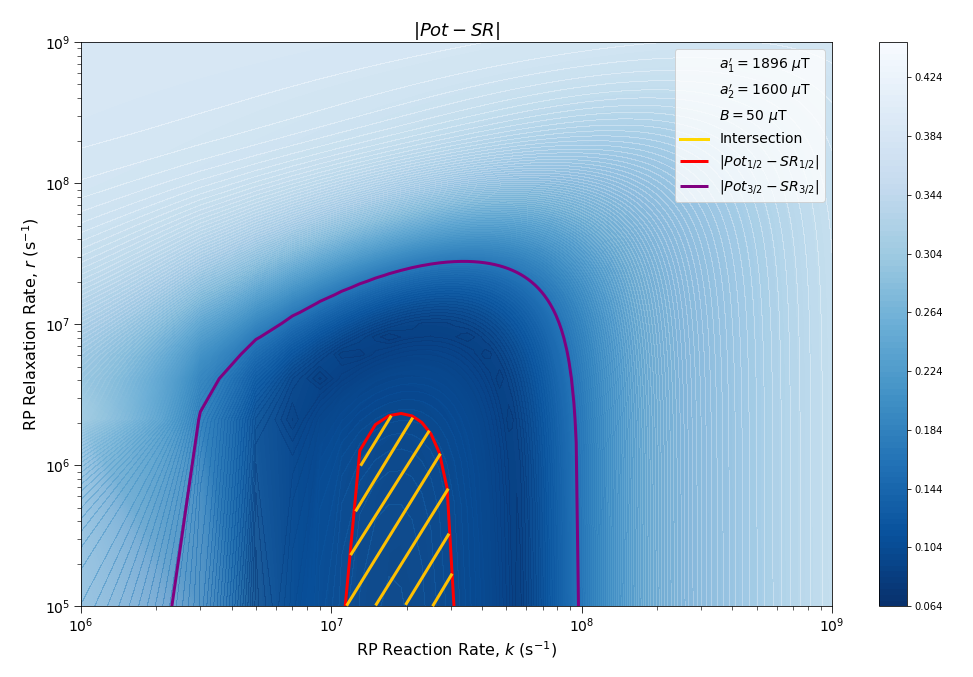}}
\caption{The dependence of the agreement between relative anesthetic potency and singlet yield ratio on $r$ and $k$ for the case where the dominant HFI for electron B is assumed to be with the TrpH$^{\cdot +}$ $\beta$-hydrogen instead of nitrogen. Here $a_1' = 1896\ \mu$T, $a_2' = 1600\ \mu$T, and $B = 50\ \mu$T. The singlet yield ratio $SR$ is calculated using $a_1 = (\gamma_n / |\gamma_{^{129}\mbox{Xe}}|) \gamma_e a_1'$, where the magnitude of the isotopic nuclear gyromagnetic correction factor, $(\gamma_n / |\gamma_{^{129}\mbox{Xe}}|)$, is normalized for $^{129}$Xe. The model can explain the experimentally derived \cite{xenon_mice} relative anesthetic potency of xenon where $\lvert Pot_{1/2} - SR_{1/2}\rvert \leq 0.08$ and $\lvert Pot_{3/2} - SR_{3/2}\rvert \leq 0.08$ intersect, shaded in yellow.}
\label{xenon_betahydrogen_rk}
\end{figure}

 One potential challenge for the specific scenario involving O$_2^{\cdot -}$ are the requirements on the spin relaxation rate $r$. Hogben \textit{et al.} \cite{hogben2009} and Player and Hore \cite{player2019} have discussed a radical pair involving O$_2^{\cdot -}$ in the context of magnetoreception. These authors argue that for free O$_2^{\cdot -}$ in solution the spin relaxation lifetime is expected to be below 1 ns due to molecular rotation. The relaxation rate requirements that we find from our calculations come within about an order of magnitude of this value, as shown in Figs. \ref{hyper_r_dep1}/\ref{hyper_r_dep_comb1} and Figs. \ref{rk_dep1}/\ref{rk_dep1_comb}. The gap widens to about two orders of magnitude if one considers $\beta$-hydrogen as the dominant nuclear spin for the electron on Trp instead of nitrogen, as shown in Fig. \ref{xenon_betahydrogen_rk}. However, the same authors have also pointed out that the oxygen spin relaxation rate could be reduced if the biological environment reduces the molecular symmetry and inhibits molecular rotation. Moreover Kattnig \cite{kattnig2017} has shown that the involvement of a paramagnetic scavenger can significantly reduce the constraints on the spin relaxation rate in the context of magnetoreception, which suggests the possibility that the same principle may apply to isotope effects in the present context. Alternatively, other RPs not including O$_2^{\cdot -}$ might also be able to explain the isotope effects. 
 


To gain a deeper understanding of the RP formation and recombination dynamics within our proposed scenario, it would be of interest to perform DFT modelling of the electron transfer between ROS and Trp with and without the presence of xenon. Such modelling could be expanded on by exploring the molecular dynamics of the binding site using quantum mechanics/molecular mechanics (QM/MM) simulation techniques, similar to those used in the case of cryptochrome \cite{mendive2018multidimensional}, to more clearly understand the stability of the radicals and other characteristics of the currently proposed RP system. In particular it would be of interest to attempt to theoretically reproduce the experimental ESR results of Turin {\it et al.} \cite{xenon_fly}. 

\begin{figure}[ht!]
\centerline{
\includegraphics[width=1.10\textwidth,height=1.0\textheight,keepaspectratio]
{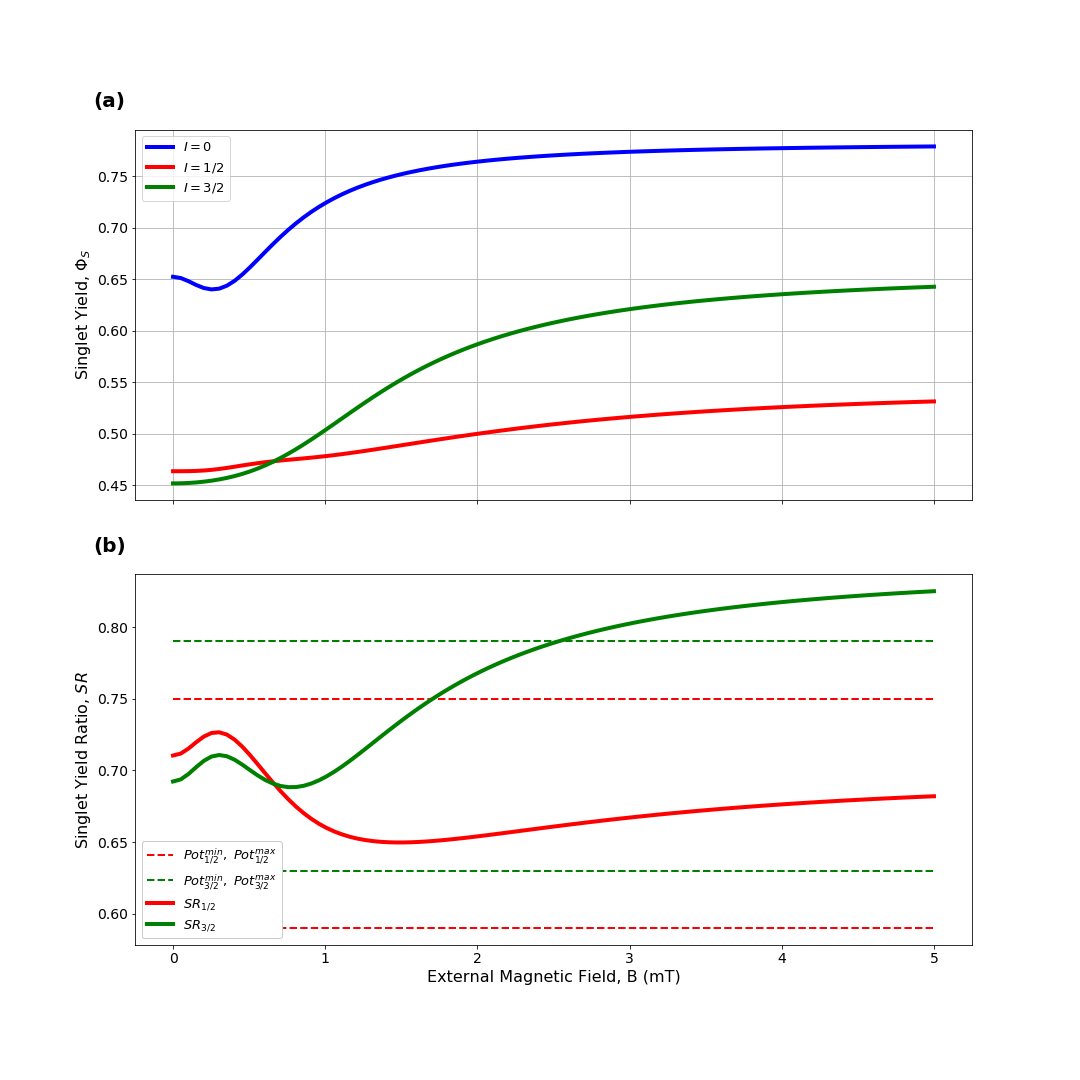}}
\caption{The dependence of the singlet yield and singlet yield ratio on an external magnetic field, with $B \in [0,\ 5000]\ \mu$T, $k = 5.5 \times 10^{7}$ s$^{-1}$, $r = 5.0 \times 10^6$ s$^{-1}$, and hyperfine constants of $a_1' = 1896\ \mu$T, $a_2' = 347\ \mu$T. The singlet yield and the singlet yield ratio $SR$ are calculated using $a_1 = (\gamma_n / |\gamma_{^{129}\mbox{Xe}}|) \gamma_e a_1'$, where the magnitude of the isotopic nuclear gyromagnetic correction factor, $(\gamma_n / |\gamma_{^{129}\mbox{Xe}}|)$, is normalized for $^{129}$Xe. \textbf{(a)} The absolute singlet yield using xenon isotopic nuclear spin values of $I \in \{0,\ 1/2,\ 3/2\}$. \textbf{(b)} The relative singlet yield ratios $SR_{1/2}$ and $SR_{3/2}$ with the singlet yield ratio of spin $I=0$ being normalized to $SR_0=1$. Values of $SR$ and $Pot$ agree for $B \in [0,\ 2525]\ \mu$T.}
\label{xenon_plot_extfield}
\end{figure}

When considering the sensitivity of the model to changes in external magnetic field, as shown in Fig. \ref{xenon_plot_extfield}, the range of $B$ values that produce agreement between $SR$ and $Pot$ is given as $B \in [0,\ 2525]\ \mu$T, and includes the geomagnetic field at different geographic locations (25 to 65 $\mu$T) \cite{field}. This result indicates that for external field values vastly larger (by approximately 2 orders of magnitude) than the geomagnetic field, the anesthetic potency of xenon may be significantly different than that observed by Li \textit{et al.} \cite{xenon_mice}; specifically, the potency of $^{131}$Xe may be greater than experimentally observed. It would be of interest to investigate the experimental effects of the external magnetic field strength on xenon-induced general anesthesia \textit{in vivo}. For example, such an experiment could involve the measurement of anesthetic potency of xenon isotopes with various nuclear spin values, including both zero spin and non-zero spin, in an environment with controllable external magnetic field.

In conclusion, our results suggest that xenon-induced general anesthesia may fall within the realm of quantum biology, and be similar in nature to the proposed mechanism of magnetoreception involving the cryptochrome protein \cite{hore2016radical}. 

This also raises the question whether the action of other anesthetics involves similar mechanisms to the one proposed here. It would be interesting to explore isotopic nuclear-spin effects as well as magnetic field effects in experiments with other general anesthetic agents that are thought to function similarly to xenon, such as nitrous oxide and ketamine \cite{gases,gases2}. 

General anesthesia is clearly related to consciousness, and it has been proposed that consciousness (and other aspects of cognition) could be related to large-scale entanglement \cite{adams2020quantum,fisher2015quantum,conscious,conscious3}. RPs are entangled and could be a key element in the creation of such large-scale entanglement, especially when combined with the suggested ability of axons to serve as waveguides for photons \cite{kumar2016possible}. Viewed in this - admittedly highly speculative - context, the results of the present study are consistent with the idea that general anesthetic agents, such as xenon, could interfere with this large-scale entanglement process, and thus with consciousness.  

\section*{\label{sec:Meth}Methods}


\subsection*{\label{subsec:DFT}DFT Analysis}

We used the Gaussian package \cite{g16} with B3LYP functional and TZVP basis set for our DFT calculation of $a_{2, exp}'$.

The range of values calculated for $a_{1, exp}'$ using DFT analysis varied widely ($a_{1, exp}' \in [955,\ 2644]\ \mu$T), depending on the functional and basis set used. The calculation is also very sensitive to the separation between O$_2^{\cdot -}$ and Xe, and the potential energy is minimized for a separation of $\sim 3$ \AA. Ultimately, the ORCA package \cite{neese2012} was used for our Xe-O$_2^{\cdot -}$ DFT calculations, and the molecular structure was optimized using the dispersion-corrected PBE0 functional and def2-TZVP basis set. The resulting calculation gives $a_{1, exp}' = 1767\ \mu$T, which is close to the value of $a_1'$ for which our model minimizes the quantity $CE = (\lvert Pot_{3/2} - SR_{3/2} \rvert + \lvert Pot_{1/2} - SR_{1/2} \rvert)$.

The orbitals obtained from the molecular structure optimization were used to calculate orbital energies as well as the hyperfine coupling constant $a_{1, exp}'$, using the B3LYP functional and def2-TZVP basis set. Relativistic effects were treated by a scalar relativistic Hamiltonian using the zeroth-order regular approximation (ZORA). Solvent effects were included with the conductor-like polarizable continuum model (CPCM), using a dielectric constant of 2. 

\begin{figure}[ht!]
\centerline{
\includegraphics[width=0.5\textwidth,height=1.0\textheight,keepaspectratio]
{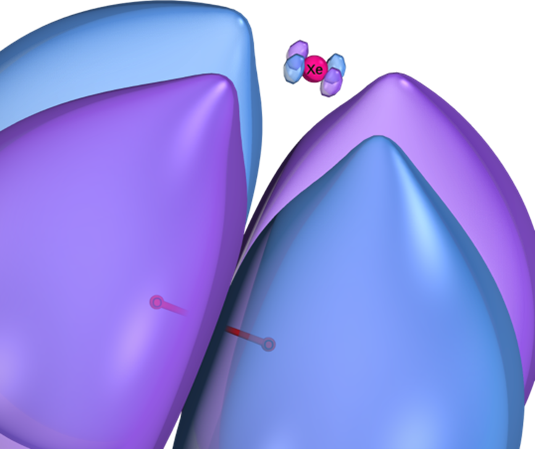}}
\caption{The highest occupied molecular orbital (HOMO) of Xe-O$_2^{\cdot -}$. Image generated using IboView v20150427 (http://www.iboview.org).}
\label{dft_homo}
\end{figure}

Our DFT computations show that the unpaired Xe-O$_2^{\cdot -}$ electron is bound, and that the highest occupied molecular orbital (HOMO) resides primarily on the O$_2$ molecule but is extended slightly onto the xenon atom, as shown in Fig. \ref{dft_homo}.

Isotropic hyperfine coupling is a result of the Fermi contact interaction between a nucleus and electron \cite{bucher2000electron}, and here this is a measure of the contribution of the valence \textit{s}-orbital to the molecular orbital occupied by the unpaired radical electron. If the unpaired electron resides mainly in a \textit{p}- or a \textit{d}-orbital, the Fermi contact interaction can also be significant where the spin density at the nucleus is induced by configuration interaction or spin polarization \cite{karunakaran2018spin}. In our case, the unpaired electron mainly occupies the \textit{p}-orbital which results in a significant isotropic HFI. The isotropic hyperfine coupling constant can be calculated as \cite{improta2004interplay, atkins2011molecular}

\begin{equation}
a_{j} = - \frac{2}{3} g_e \gamma_e \gamma_n \mu_0 |\Psi(0)|^2
\label{HFCC}
\end{equation}

\noindent where $g_e$ is the electron g-factor, $\mu_0$ is the permeability of free space, and $|\Psi(0)|^2$ is the electron probability density evaluated at the nucleus.

\section*{Data Availability}

The datasets generated and analysed during the current study are available from the corresponding author on reasonable request.

\bibliography{ref_file}

\section*{\label{sec:Ack}Acknowledgements}

The authors would like to thank Peter Hore, Robert Dickinson, Sourabh Kumar, Parisa Zarkeshian, Rishabh, Sumit Goswami, Faezeh Kimiaee Asadi, Stephen Wein, Jiawei Ji, Yufeng Wu, Kenneth Sharman, Rogelio Delgado Venegas, Jiri Hostas, Morteza Chehel Amirani, Lizandra Barrios Herrera, and Sridhar Dwadasi for their input, comments, and insights. C.S. particularly thanks Stuart Hameroff for bringing the results of Li \textit{et al.} \cite{xenon_mice} to his attention. The authors would like to acknowledge Compute Canada for its computing resources. This work was supported by the Natural Sciences and Engineering Research Council of Canada.

\section*{\label{sec:cont}Author Contributions Statement}
J.S. performed calculations and modelling with help from H.Z.H. and C.S.; H.Z.H. performed DFT analysis with help from J.S., D.S., and C.S.; J.S. and C.S. wrote the paper with feedback from D.S. and H.Z.H.; C.S. conceived and supervised the project. 

\section*{Competing Interests}

The authors declare no competing interests.

\end{document}